\input epsf
\documentclass[nohyper,notoc]{article} 
\usepackage{epsfig}
\usepackage{amsmath} 

\def\bit{\begin{itemize}}
\def\eit{\end{itemize}}
\def\ben{\begin{enumerate}}
\def\een{\end{enumerate}}
\def\beq{\begin{equation}}
\def\eeq{\end{equation}}
\def\bea{\begin{eqnarray}}
\def\eea{\end{eqnarray}}
\def\bq{\begin{quote}}
\def\eq{\end{quote}}
\newcommand{\bma}{\begin{pmatrix}}
\newcommand{\ema}{\end{pmatrix}}
\def \lsim{\mathrel{\vcenter
     {\hbox{$<$}\nointerlineskip\hbox{$\sim$}}}}
\def \gsim{\mathrel{\vcenter
     {\hbox{$>$}\nointerlineskip\hbox{$\sim$}}}}
\def\gappeq{\mathrel{\rlap {\raise.5ex\hbox{$>$}}
{\lower.5ex\hbox{$\sim$}}}}
\def\lappeq{\mathrel{\rlap{\raise.5ex\hbox{$<$}}
{\lower.5ex\hbox{$\sim$}}}}

\def\dslash{ \, \partial  \! \! \! \! / ~ }

\def\meg{\mu \to e \gamma}
\def\teg{\tau \to e \gamma}
\def\tlg{\tau \to \ell \gamma}
\def\tmg{\tau \to \mu \gamma}
\def\llg{\ell_\a \to \ell_\b \gamma}
\def\m3e{\mu \to e \bar{e} e}
\def\a{\alpha}
\def\b{\beta}

\def\m{\mu}

\begin{document}
\renewcommand{\thefootnote}{\fnsymbol{footnote}}
\begin{center}
{\Large {\bf 
Reconstructing Seesaws}}
\vskip 25pt
{\bf   Sacha Davidson $^{2,}$\footnote{E-mail address:
s.davidson@ipnl.in2p3.fr} and  Martin Elmer $^{3,}$}\footnote{E-mail address:
m.elmer@ipnl.in2p3.fr} 
 
\vskip 10pt  
$^1${\it } \\
$^2${\it IPNL, Universit\'e de Lyon, Universit\'e Lyon 1, CNRS/IN2P3, 
4 rue E. Fermi 69622 Villeurbanne cedex, France}\\
\vskip 20pt
{\bf Abstract}
\end{center}

\begin{quotation}
  {\noindent\small 
We explore some aspects  of ``reconstructing'' the 
heavy singlet sector 
of  supersymmetric type I 
seesaw models, for two, three or four singlets. We 
work in the limit where one light neutrino is massless.  In an
ideal world, where selected coefficients of the TeV-scale
effective Lagrangian could be measured with arbitrary
accuracy, the two-singlet case can be reconstructed,
two three or more singlets can be differentiated,
and an inverse seesaw with four singlets can be
reconstructed. In a more realistic world, we estimate
$\ell_\a \to \ell_\b \gamma$  expectations with a 
``Minimal-Flavour-Violation-like'' ansatz,
which gives a relation between  ratios of the three branching ratios.
The two singlet model predicts a discrete set of
ratios. 
\vskip 10pt
\noindent
}

\end{quotation}

\vskip 20pt  

\setcounter{footnote}{0}
\renewcommand{\thefootnote}{\arabic{footnote}}


\section{Introduction}
\label{intro}
New particles which are too heavy to be
produced on-shell, can nonetheless leave observable
traces of their virtual exchange. At scales well below
their mass, their effects can be described by an 
effective Lagrangian containing non-renormalisable
operators induced by the exchange of the heavy new particles. 
While it is clear that the effective
Lagrangian can always be constructed 
from a New Physics model, the prospects for
``reconstructing'' the New Physics from
the effective Lagrangian are nebulous. 
Such a ``reconstruction'' would be 
interesting for any New Physics
 at a scale beyond the reach of the LHC.

This paper aims  to analytically
explore this reconstruction in the
simplest of toy models.
 We suppose that light neutrino
masses, with one massless neutrino, are
generated by  a supersymmetric type I 
seesaw model\cite{seesaw},  with  two,
three or four heavy singlets of mass
$M_I \gg $ TeV \footnote{This
means we do not consider ``low-scale'' seesaws,
with singlet masses in the eV $\to$ TeV range
\cite{lowscale}.}. 
When the number
of singlets $n_{N}$ does not exceed
the number of lepton doublets,
this model is in principle ``reconstructable''  \cite{DI1},
that is,  the masses and
mixing angles of the singlet sector
can be computed from parameters 
(masses and mixing angles)
of the doublet sector. This
reconstruction has been studied 
from numerous perspectives \cite{DIP,2RHN,snus,CIJA,IMP,Federica}.
Analytic formulae for   the reconstruction
of  two-singlet models  were presented in  \cite{2RHN}.
  We wish to know what  we could
learn about the singlets, in principle
and also with some degree of realism,  from effective operators
involving doublets. So  we start in section
\ref{2N}  with a model containing
two singlets, which has sufficiently
few parameters that it could be disfavoured
by observations. Then in section \ref{3N} we consider  models with  three
singlets, in two limits: where there is
a massless doublet neutrino because
a Yukawas eigenvalue vanishes, and
the case where a singlet is infinitely
massive. As expected,  the three-singlet sector is in principle
reconstructable when there are three non-zero Yukawa
eigenvalues, 
but not when there are only two, 
 because in this latter case  the singlet components
which couple via the zero Yukawa are
decoupled.  Finally, in section \ref{4N}, by
considering even more exotic
operators, we show that in principle
it could be possible to distinguish 
between  two, three or more singlets models,
and to reconstruct a four singlet
 ``inverse seesaw'' \cite{inverse} model.


\section{ Notation and Assumptions}
\label{}

 The seesaw model \cite{seesaw}  is a natural and minimal
extension of the Standard Model which  fits the observed
neutrino masses. We consider a supersymmetric seesaw model
(with conservation of R-parity\footnote{We assume 
that neutrino masses arise due to the seesaw, so
R-parity, or some other symmetry, must prevent
\cite{goran} other
dangerous lepton number violating interactions
which could generate majorana neutrino masses
\cite{RPmnu}.})
because  we are interested in seesaw reconstruction;
 the slepton  masses may contain additional
information about the seesaw parameters \cite{DI1}. 
Also, in the non-supersymmetric seesaw, there are
${\cal O} ((y^\nu M)^2)$ contributions to
the Higgs mass which must be fine-tuned. 

At scales above the singlet masses $M_I$, the
superpotential can be written
\beq
W_{lep} =
  [{\bf Y^e}]_{\a\a} (L_\a H_d ) E^c_\a +
[ {\bf Y^\nu}]_{\b J} (L_\b H_u)  N_J^c + \frac{M_I}{2} N_I^c N_I^c.
\label{W}
\eeq 
This expression is in the eigenbases of ${\bf Y^e} {\bf Y^e}^\dagger$
for the $\{ L_\a\}$,
 ${\bf Y^e}^\dagger {\bf Y^e}$ 
 for   the  $\{ E^c_\b \}$, 
   and ${\bf M}$ for the
$\{N_I \}$.
These choices correspond to  the charged lepton
mass basis, refered to as the flavour
basis  and  labelled by
greek letters, and 
the mass eigenstate basis of
the heavy singlets, labelled by 
roman capitals which run from
1 to $n_N$.
The doublet contraction
is antisymmetric  $(L H_d) =
E_L H^0 - N_L H^-$, the Yukawa indices are   
ordered left-right,
and we will allow  $n_{N} = 2,3,4$  generations 
 of singlets  $N$, whose  masses $M_I$ will
usually be taken $ \gg$ TeV.
The resulting  Lagrangian  is 
\beq
{\cal L} =   y^e_{{\alpha}} \overline{\ell}^{\alpha}  H^*_d  e_R^{\alpha}+ 
   \overline{\ell}^{\alpha}   H^*_u  [  {\bf Y^\nu}^*]_{\alpha I} N_I  + 
 \frac{{M}_I}{2}  \overline{N^c}_I N_I +...+ h.c.
\label{L}
\eeq
where $y^e_\a  \in \{ y_e, y_\mu, y_\tau \}$
 are the charged lepton Yukawa couplings,
 the singlet neutrinos are written as
four-component fermions, and 
the  ... includes   sparticle
interactions.

The number of parameters \cite{paramcount} in the  superpotential
(\ref{W}), or Lagrangian (\ref{L}), will depend on
the number of singlets $n_N$, and on whether the
elements of the matrices {\boldmath $M$} and {\boldmath ${\bf Y^\nu}$ }
are allowed to be complex or restricted to be real.
  In the case of complex matrices with  $n_N$
singlets, there are $n_N$  masses $\{ M_I \}$,
which can be taken real by a phase choice on
the $\{ N_I\}$,  three eigenvalues $\{ y^e_\a \}$,
which can be taken real by a relative  phase choice 
between 
the $\{ E^c_{\a}\}$ and  $\{ L_{\a}\}$, and  $3 \times n_N$ complex entries
in ${\bf Y^\nu}$, from which 3 phases can be
removed by suitably choosing the  phase differences 
between  the  three doublets  
 $\{ L_{\a}\}$ and the singlets  $\{ N_I\}$. So we expect $7 \times n_N$ 
real parameters.
 If the  matrices 
{\boldmath $M$} and {\boldmath ${\bf Y^\nu}$ }
are restricted to be real, 
there would be 3 + $4 \times n_N$ 
 real parameters.

 The neutrino Yukawa matrix ${\bf Y^\nu}$ is a
$3 \times n_{N}$ matrix, with  at most 
$min \{ 3,n_N \}$  non-zero eigenvalues.
It  can be diagonalised
by independent unitary transformations 
 ${\bf V}_L$  (which acts in on
the left and is $3 \times 3$) 
and ${\bf V}_R$(which is  $n_N \times n_N$
and acts on  the right). So in the case
of $n_N = 3$ :
\bea \label{yuk}
 {\bf V}_L {\bf Y^\nu} 
 {\bf V}_R^{\dagger} &= &D_{Y^\nu}\equiv {\rm diag} ~ \{ y_1, y_2,  y_3 \} 
\eea
For $n_N = 2$,   the same formula
can be used, but  with $y_1 = 0$, 
and by putting the  non-trivial $2 \times 2$ 
$V_R$  in the lower right corner of the 
$3 \times 3 $ matrix appearing in eqn (\ref{yuk}):
\beq
[{\bf V}_{R}] _{3 \times 3} = 
\left[
\begin{array}{lc}
1 & 0 ~~ 0 \\
0 &  [{\bf V}_R]_{2 \times 2}\\
0 &  \\
\end{array}
\right]
\eeq 
The case where $n_N = 4$, can be similiarly delt with,
by embedding the $3 \times 3$ $V_L$ into a $4 \times 4$
matrix.


At scales $\ll M$, where the 
 $N$ are not present as on-shell
particles, the effective Lagrangian will
contain non-renormalisable operators induced
by $N$ exchange. For simplicity, in this
paper we focus on CP conserving observables
(for instance, we neglect electric dipole moments
\footnote{For a review and references, see
{\it e.g.} \cite{ybook}}).

At dimension five,  arises the majorana
mass operator for the doublet  neutrinos
\beq \label{m_nu_eff}
{\bf m_{\nu}} = {\bf {\bf Y^\nu}} {\bf M}^{-1} {\bf Y^\nu}^{T} v_u^2 
= U {\bf D_{m}} U^T~~,
\eeq  
where  $v_u = \langle H_u \rangle = v \sin \beta \simeq v = 174$ GeV,
 $D_{m} = diag \{ m_1, m_2 ,m_3\}$, and the neutrino
mass differences are taken to be
\beq
\label{m2}
|\Delta m^2_{31}| \simeq 2.4 \times 10^{-3} ~ {\rm eV}^2
~~~~
|\Delta m^2_{21}| \simeq 7.6 \times 10^{-5} ~ {\rm eV}^2
\eeq
We assume that  the lightest doublet
neutrino is exactly massless  (relaxing this
assumption is discussed at the end of 
section \ref{M3}), and consider
separately the hierarchical(NH: $m_3>m_2>m_1 = 0$))  and 
inverse hierarchical (IH: $m_2>m_1>m_3 = 0$) patterns. 
 $U$ is the  leptonic mixing   matrix
\bea \label{U_PMNS}
    {\bf U} &=&\left[
    \begin{array}{ccc}
         e^{i \alpha}~ c_{13} c_{12}
        &e^{i \beta}~ s_{12} c_{13}
        & s_{13}~ e^{-i \delta}  \\
         e^{i \alpha}~(-s_{12} c_{23} - s_{23} s_{13} c_{12}~e^{i \delta})  
        &  e^{i \beta}~(c_{23} c_{12} - s_{23} s_{13} s_{12}~ e^{i \delta}) 
        &  s_{23} c_{13} \\
        e^{i \alpha}~ (s_{23} s_{12} - s_{13} c_{23} c_{12}~e^{i \delta})
       &  e^{i \beta}~(-s_{23} c_{12} - s_{13} s_{12} c_{23}~e^{i \delta})
        &  c_{23} c_{13}
    \end{array} \right]
\eea
where $c_{ij} = \cos \theta_{ij}$, etc.,
 $\theta_{23} = \pi/4$\cite{PDG}, 
$\sin^2 \theta_{12} = 0.3$  \cite{PDG}
(or sometimes $\sin^2 \theta_{12} = 1/3$),
 and 
$U_{e3}  =0.18 $,
the central
value of the recent T2K indication\cite{T2K}   of
non-zero $\theta_{13}$ (averaged over
NH  and IH).  Since  a neutrino is  massless,
one of the Majorana phases $\alpha , \beta$
vanishes. In practise, we will usually neglect
all phases for simplicity (we do not discuss CP
violating observables).

  At
dimension six, the seesaw model induces
the lepton number conserving
operator  \cite{dGGST,BGJ}
$(\overline{\ell_\a} H^*) \dslash (\ell_\b H)$
with coefficient 
\beq
 [{\bf d}]_{\a \b} 
 \equiv 
\left[{\bf Y^\nu}{\bf M}^{-1} {\bf M}^{-1 *} {\bf Y^\nu}^\dagger 
\right]_{\a \b} ~~.
\label{opBGJ}
\eeq
 After
electroweak symmetry breaking, this operator
modifies the neutrino kinetic terms,
so can contribute to the  non-unitarity
of the leptonic mixing  matrix\cite{non-U}.

The third operator which we wish to use for
 seesaw reconstruction is the  dimension
six  electromagnetic
dipole operator, which  can be written, after
electroweak symmetry breaking, as
\beq
g \sin \theta_W {\Big (}
 [{\bf X}_{L}]_{\a \b} \overline{e_\a} \sigma ^{\mu \nu} P_L e_\b F_{\mu \nu}
+
 [{\bf X}_{R}]_{\a \b} \overline{e_\a} \sigma ^{\mu \nu} P_R e_\b F_{\mu \nu} 
 {\Big )}
~~~.
\label{dipole}
\eeq
We assume that the coefficients  $[{\bf X}_{L}]_{\a \b}$ of this 
chirality-flipping operator have the 
``Minimal Flavour Violation\cite{dAGIS}-like'' 
form 
\beq
\label{Xmfv}
{\bf X}_L  =  v  C^X_{e} {\bf Y^e}^\dagger
+ v C^X_{\nu}  {\bf Y^e}^\dagger {\bf Y^\nu} {\bf Y^\nu}^\dagger ~~~,
\eeq
where $v = 174 $ GeV is the Higgs
vev, and the   
 coefficients $C^X_x$ have mass
dimension -2 and are assumed 
flavour independent.
Information 
about  the  flavour off-diagonal  
elements of $[{\bf Y^\nu} {\bf Y^\nu}^\dagger]$
therefore can be obtained.
This form 
 is ``MFV-like'',
because the lowest order dimensionless lepton flavour-changing
operator in the  seesaw model is  
$[{\bf Y^\nu} {\bf Y^\nu}^\dagger]$.
  This  parametrisation of $X_{L}$
  is motivated by supersymmetry,
where loop corrections to the soft masses can 
approximately give
such a dependance on    $[{\bf Y^\nu} {\bf Y^\nu}^\dagger]$.
In the Appendix,
we review the supersymmetric motivation
for eqn (\ref{Xmfv}), and estimates for  the
coefficients  $C^X_\nu $ and  $C^X_e $.
These imply that 
\bea
\label{small}
 [{\bf Y^\nu} {\bf Y^\nu}^\dagger]_{\mu e}
&  \lsim &  2 \times 10^{-3} ~~~,
\eea
to suppress the rate for $\meg$. 
We impose this ``approximate zero'' on our models,
because it  approximately fixes
a parameter, with a weak dependence on the
value of $C_\nu^X$.   
 We try to avoid assuming  
 a value for $C_\nu^X$,
because  a precise calculation 
depends on several  model parameters at various
scales.  This means we
extract parameters from ---or predict --- ratios
of  lepton flavour violating rates (such
as eqn (\ref{rapport})), but we
cannot predict the rates (such
as eqn (\ref{radiative_decay})).

Lepton flavour violating 
radiative
decays,  $\ell_\a \to \ell_\b \gamma$,
  proceed
via the operator 
(\ref{dipole})
at a rate given by 
\bea \label{radiative_decay}
\widetilde{BR} (l_{\alpha} \to l_{\beta} ~ \gamma)
\equiv \frac{\Gamma (l_{\alpha} \to l_{\beta} ~ \gamma)}
{\Gamma(\ell_\alpha \rightarrow \ell_\beta \nu_{\alpha} \bar{\nu}_{\beta})} 
&   =  & 
\alpha m_\alpha^3 (|X_{L \a \b}|^2 + |X_{R \a \b}|^2) \frac{192 \pi^3} 
{G_F^2 m_\alpha^5} ~~.
\eea
Table \ref{tab} lists the current bounds,  and hoped
for sensitivities of running or planned experiments.
  \begin{table}
\renewcommand{\arraystretch}{1.25}
$$\begin{array}{|c|c|c|}\hline 
\widetilde{BR} & \hbox{current ~ bound} & \hbox{future} \\
\hline
\meg & 2.4 \times 10^{-12}\cite{MEG} & \sim 10^{-13} , \hbox{(MEG \cite{MEG})} \\
\tmg &  2.5 \times 10^{-7} \cite{tmgexp} & \sim 10^{-8} , 
\hbox{(super-B~ factories\cite{SB})} \\
\teg &  1.9 \times 10^{-7}\cite{tmgexp} & \sim 10^{-8} , 
\hbox{(super-B~ factories\cite{SB})} \\
\hline
\end{array}$$
\caption{Current bounds  and hoped-for sensitivities
to lepton flavour violating branching ratios,  normalised
to leptonic weak decays, as 
in eqn (\ref{radiative_decay}). 
\label{tab}}
\renewcommand{\arraystretch}{1.00}
\end{table}
If two different lepton flavour violating decay rates were
observed,  for instance $\meg$ and 
$\tmg$, then the approximation (\ref{Xmfv})  implies 
the following equality
\beq
\label{rapport}
\frac{\widetilde{BR}(\meg)}{\widetilde{BR}(\tmg)} =
\frac{|[{\bf Y^\nu} {\bf Y^\nu}^\dagger]_{\mu e}|^2}
{|[{\bf Y^\nu} {\bf Y^\nu}^\dagger]_{\tau \mu }|^2} ~~.
\eeq
We  assume that  the eigenvalues of 
${\bf Y^\nu} {\bf Y^\nu}^\dagger = V_L^\dagger D_\nu^2 V_L$
 are sufficiently  hierarchical, that in  $\ell_\a \to \ell_\b \gamma
$ processes, 
 only
the terms  involving the largest eigenvalue  $ y_3$
need to  be considered:
\beq
\label{approxhier}
|[{\bf Y^\nu} {\bf Y^\nu}^\dagger]_{\a \b}|^2
\longrightarrow
y_3^4|V_{L 3\a} V^*_{L 3 \b}|^2
 ~~.
\eeq
 As all known Yukawa eigenvalues of the SM are hierarchical, 
this assumption is not unreasonable for the neutrino Yukawa matrix.
The  approximation (\ref{approxhier}) implies a 
relation among the three  rates 
$\ell_\a \to \ell_\b \gamma$, because
they are all controlled by one row of
$V_L$, which depends on  two angles.
The $y_3^4$ in eqn (\ref{approxhier}) 
cancels in {\it ratios} of
branching ratios such as  eqn(\ref{rapport}).
The $\llg$ branching ratios are $\propto y_3^4$
(an allowed range for $y_3$ might be disentangled
from slepton masses, if the spectrum of
supersymmetric particles was well-known); 
we   set  $y_3^4 = 1$ to estimate
branching ratios.  
If  the assumption of sufficiently hierarchical 
Yukawa eigenvalues is dropped, the  $\ell_\a \to \ell_\b \gamma$ 
rates depend on additional parameters. 
This destroys the relations between different 
rates.

The aim of this study, is to ``reconstruct''
the singlet sector of seesaw models.  With
this aim, we
consider three matrices in doublet
flavour space: the light neutrino
mass matrix  $[{\bf m}_\nu]$
of eqn (\ref{m_nu_eff}),
the dimension six
 \cite{BGJ}
 operator 
$[{\bf d}]$
of eqn (\ref{opBGJ}),
and  the matrix
\bea
{\bf P}  &=& {\bf Y^\nu} {\bf Y^\nu}^\dagger \equiv V_L^\dagger {\bf D}^2_{\nu} V_L 
\label{P}
\eea
The first  two are  coefficients of operators in the effective
lagrangian, and the  off-diagonal elements of ${\bf P}$  
 could be extracted from  the slepton mass matrix.
To  finish determining ${\bf Y^\nu} {\bf Y^\nu}^\dagger$
(in the approximation of  neglecting phases),
three more inputs would be required, which we take to
be the eigenvalues. We  assume  they
are hierarchical,  that 
$y_3 = 1$, and we will keep
the remaining two as free parameters.
We do not study the  extraction of  these
eigenvalues from slepton mass matrices, although
this can be envisaged \cite{snus,2RHN} in some
circumstances.

If the elements  of  the light  neutrino mass
matrix, and  of either of ${\bf P}$ or $[{\bf d}]$  could
be measured exactly,  
then the parameters of the high scale Lagrangian 
of eqn (\ref{L}) can be reconstruct  in the $n_N \leq 3$ seesaw.
This is clear in the case of  ${\bf P}$
and $[{\bf m}_\nu]$:
diagonalising  $ P = {\bf Y^\nu} {\bf Y^\nu}^\dagger$
gives the  ${\bf Y^\nu}$ eigenvalues 
and the unitary matrix $V_L$ which transforms from
the doublet basis where   ${\bf Y^e}$ is diagonal,
to the basis where  ${\bf Y^\nu}$ is diagonal. 
The matrix
\beq
W_L = V_L U ~~~,
\label{WL}
\eeq
which  tranforms from the neutrino
mass basis to the basis  ${\bf Y^\nu}$  diagonal,
is therefore known.  So the expression
for the  mass matrix of massive light neutrinos 
\bea \label{Dnu}
{\bf D}_m  & = & { W_L}^\dagger {\bf D_{\nu}} V_R{\bf D_M}^{-1}{ V_R}^T 
{\bf D}_{\nu}
{ W}_L^*   v_u^2 
\eea
can be inverted for  $V_R$ and $D_M^{-1} = {\rm diag}
\{ 1/M_1, ... \} $:
\bea \label{DM-1}
  V_R  {\bf D_M}^{-1}{ V_R}^T 
 & = &   {\bf D_{\nu}}^{-1} { W_L} {\bf D}_m 
{ W}_L^T {\bf D}_{\nu}^{-1}    v_u^2 
\eea
provided that ${\bf D}_\nu$ has no zero eigenvalues.
 The matrix $W_L$  parametrises
the deviation of $V_L$ (whose angles control
LFV processes)  from the leptonic mixing matrix $U$. 
If  there are only two non-zero eigenvalues in 
 ${\bf Y^\nu}$, so the lightest
neutrino is massless,  
then $W_L$ is a $2 \times 2$ matrix,
and $V_L$ differs from $U$ by a single rotation.

As discussed by Broncano,Gavela and Jenkins\cite{BGJ},
the parameters of the high-scale singlet lagrangian
can also be reconstructed  from 
 $[{\bf m}_\nu]$ and  $[{\bf d}]$.
Although $[{\bf d}]$  is the coefficient of a 
dimension six operator, (so  suppressed $\propto 1/M^2$),
if it was
known exactly, then
\beq
v^4 [{\bf m_\nu}]^{-1 } [{\bf d}] \, [ {\bf m_\nu}^*]^{-1 } = 
[{\bf Y^\nu}^{T}]^{ -1}  [{\bf Y^\nu}]^{* -1} 
\eeq
so the matrices
 $[{\bf Y^\nu}]$ and  $[{\bf M}]$
can be obtained. 

In an ideal world, where two of the  the three matrices
 ${\bf P}$,  $[{\bf m}_\nu]$ and $[{\bf d}]$ could
be measured, then the third could be used as a test of
the seesaw.

In summary, we assume that the lightest doublet
neutrino is massless, that $\theta_{13} \simeq .18$,
and we neglect the phases of the lepton mixing matrix. 
We assume that the neutrino Yukawa
coupling matrix ${\bf Y^\nu}$ has hierarchical
eigenvaluess with $y^\nu_3 = 1$, and
that  flavour off-diagonal elements
of $[{\bf Y^\nu} {\bf Y^\nu}^\dagger]_{\a \b}$ can be
extracted from  the rates
for $\ell_a \to \ell_\b \gamma$.


\section{Two singlets}
\label{2N}

The first model that we consider contains 
two heavy  singlet (``right-handed'') majorana  neutrinos $N_I$. 
This is the 
simplest see-saw model that can  explain 
all observed neutrino oscillations, and
has been carefully studied by
several people \cite{2N,2RHN}.
Here, we review the expectations for
$\llg$ decays, and the prospects for
reconstructing the singlet sector. 
This model  predicts that 
one of the active doublet  neutrinos has a  vanishing mass;
in the case of the normal hierarchy, this massless neutrino
is taken to be $\nu_1$, whereas it is $\nu_3$ for the inverse hierarchy.

In general there are two possibilities to write down this model. 
The first possibility is to use a   $3 \times 2$ 
neutrino Yukawa coupling matrix 
$[{\bf Y}^\nu]_{\a I }$, 
so that it can connect the 
3 generation doublet lepton sector to the
2 generation singlet sector. The second possibility 
is to write also the right handed sector as $ 3 \times 3 $ 
matrices and fill up the unused components with zeros, so 
that all matrices that are used can  by diagonalised. 
We will use the latter one.

\subsection{Real case}
\label{real}

In this first step, all  phases are set to zero.

To reconstruct the high scale parameters,
using  
equation (\ref{DM-1}),  we need
the eigenvalues of ${\bf Y^\nu}$, and
the matrix $W_L$, which parametrises
the deviation of $V_L$ from $U$.  
 In the two singlet model, 
equation  (\ref{DM-1})  
can be written  in terms of $2 \times 2$ matrices, 
because  the basis of the charged leptons is not used.
We take 
$y_3 = 1$,   $y_2$ remains as a
free parameter, and 
writing 
\beq
[V_L]_{3 \a}  = [W_L]_{3j}U^*_{\a j} 
\label{311}
\eeq 
allows to obtain
 the orthogonal $2 \times2$   matrix
 $W_L$  from  lepton
flavour violating decay ratios,
 as given
 in eqn
(\ref{approxhier}). 
 The
index $j$  runs from 2  to 3 for
the normal hierarchy, and 1 to 2 for
the inverse hierarchy.

The bound (\ref{small})
 from the non-observation of
 $\mu \to e \gamma$  \cite{MEG},
 combined with the
assumed hierarchy $1 = y_3 \gg y_2$
(see eqn
(\ref{approxhier})),  
approximately fixes the angle $\theta_{WL}$,
whose  value  hardly 
changes as $  |V_{L3e} V_{L3\mu}^*| $ 
varies from  $  2 \times 10^{-3}$ to zero. 
Of course this fine-tuning of an
element of $V_L$  changes 
 the branching ratios which depend on it, but has little
effect on the
reconstructed inverse  
mass matrix ${\bf M}^{-1}$ of the singlet neutrinos. 
The ``approximate zero''
 of   eqn (\ref{small}) 
can be implemented either as
 $|V_{L3e}|  = |s_{WL} U_{ej} + c_{WL} U_{ej+1} | < 10^{-2}$  
or  $|V_{L3\mu}|  = |s_{WL} U_{ \mu j} + c_{WL} U_{ \mu j+1} |   < 10^{-2}$
(where $j = 1$ for the inverse hierarchy,  $j=2$ for a normal hierarchy,
and  $s_W = \sin \theta_{WL}$, etc.). 
These two  possibilities determine the two
possible solutions for $\theta_{WL}$.
 If
 $|V_{L3e}|$ is small, then
the branching ratio for $\tmg$ could be detectable, and 
\beq
\label{VL3e0}
\frac{\widetilde{BR}(\meg)}{\widetilde{BR}(\teg)}
\simeq
\frac{
 |V_{L 3\mu}|^2}
{|V_{L 3 \tau}|^2
} 
\simeq
\frac{|-U_{ej+1}^* U_{ \mu j} + U_{ej}^* U_{ \mu j+1} |^2}
{|-U_{ej+1}^* U_{ \tau j} + U_{ej}^* U_{ \tau j+1} |^2}
 \sim {\cal O}(1)
\eeq
 Whereas if
 $|V_{L3 \mu}|$ is small, then
the branching ratio for $\teg$ could be detectable, and 
\beq
\label{VL3mu0}
\frac{\widetilde{BR}(\meg)}{\widetilde{BR}(\tmg)}
\simeq
\frac{
 |V_{L 3e}|^2}
{|V_{L 3 \tau}|^2
}  
\simeq
\frac{|-U_{ \mu j+1}^* U_{ e j} + U_{ \mu j}^* U_{ e j+1} |^2}
{|-U_{ \mu j+1}^* U_{ \tau j} + U_{ \mu j}^* U_{ \tau j+1} |^2}
 \sim {\cal O}(1)
\eeq
We now consider the expectations for these
ratios of branching ratios, for the normal
and inverse hierarchy patterns of light masses.

For the inverse hierarchy,
the case $V_{L3e} \to  0$ gives
\beq
\label{IH1}
\frac{\widetilde{BR}(\meg)}{\widetilde{BR}(\teg)}
 = 1 
\eeq
which   implies that $\teg$ is beyond the
reach of planned experiments, see table \ref{tab}. 
However,  $\tmg \propto |V_{3\mu}V_{3\tau}^*|^2 \simeq .25$,
so  according to  the SUSY estimate
for   $C_\nu^X$ (see eqn (\ref{raddec3})), 
$\widetilde{BR}(\tmg) \sim 2.3 \times 10^{-7}$,
so  should be seen
in upcoming experiments (see table \ref{tab}).

The case $V_{L3 \mu} \to  0$  in
the inverse hierarchy is curious, because 
$ V_{L3 \tau} \to  s_{13}$  and $V_{L3e} \to 1$. 
Therefore $\tmg$ is beyond the reach
of planned experiments :
\beq
\label{IH2}
\frac{\widetilde{BR}(\tmg)}{\widetilde{BR}(\meg)} \simeq
s_{13}^2 ~~,
\eeq
 and since 
$| V_{L3 \tau} V_{L3e}^* | \simeq s_{13}$,
the supersymmetric estimate of
eqn   (\ref{raddec3})
gives $\widetilde{BR} (\teg) \sim  10^{-6} s_{13}^2$.
This could be detectable,
for cooperative
supersymmetric masses and mixing angles,
and $s_{13}$ close to the current  T2K 
central value $\sim .18$. Unfortunately, with
uncooperative supersymmetric parameters,
 all $\llg$ decays could be out of reach.

Consider now the case of a normal hierarchy for the light neutrino
masses, so  the $V_{L3 \alpha}$ are defined with $j = 2..3$
in eqn (\ref{311}). 
If $|V_{L3e}| \to 0$, then
we obtain
\beq
\frac{\widetilde{BR}(\meg)}{\widetilde{BR}(\teg)} \simeq
 \frac{  |- U_{e 3} U_{\mu 2} +  U_{e2} U_{\mu 3}|^2}
{ |- U_{e 3} U_{\tau 2} +  U_{e2} U_{ \tau 3}|^2}
\simeq 0.35
\eeq
which implies that $\teg$ is beyond the reach of
planned experiments. However, the supersymmetric
estimate for $C_\nu^X$ from the Appendix implies
that  $\widetilde{BR}(\tmg) \sim  10^{-7}$. 
Alternatively, when   $|V_{L3 \mu }| \to 0$,  one  obtains
 $\widetilde{BR}(\teg) \sim 1 \times  10^{-7}$,
 and $\widetilde{BR}(\tmg)$ out of
reach at   a branching ratio of order $\widetilde{BR} (\meg)$:
\beq
\frac{\widetilde{BR}(\tmg)}{\widetilde{BR}(\meg)} \simeq
 \frac{  |- U^*_{\mu 3} U_{\tau 2} +  U^*_{ \mu 2} U_{\tau 3}|^2}
{ |- U^*_{\mu 3} U_{e 2} +  U^*_{\mu 2} U_{ e 3}|^2}
 \simeq \frac{4}{1 - 2\sqrt{2} s_{13}}
\eeq
Notice that in the normal hierarchy case,  
it is not possible to suppress all the 
$\ell_\a \to \ell_\b \gamma$ decays, as
was possible for the inverse hierarchy:
there is always  one small branching ratio of the  order of 
$\meg$, and one  $BR(\tlg)  \sim  10^{-8}$, 
which  could be measurable at a superB factory.

We can now attempt to reconstruct the mass of the heavy
singlet neutrinos, with $\theta_{WL}$ fixed by
eqn (\ref{small}). The masses are similar for
the four possible values of $\theta_{WL}$.   
The  mass $M_2$  of the lighter singlet  depends on
 $y_2^2$, where  $y_2$ is  the yet 
unfixed second neutrino Yukawa eigenvalue. 
Thermal leptogenesis \cite{PRep}
 \footnote{Our model is supersymmetric, so some solution
\cite{gravitino} to
the gravitino problem would be required. } 
prefers  $ M_2 \gsim 10^{10}~ {\rm GeV}$,
which can be obtained  for  
\bea
 0,002 \le & y_2 & \le 0,007  ~~~{\rm normal~ hierarchy} \\
  0,003 \le & y_2&  \le 0,009 ~~~
{\rm  inverse ~hierarchy} .
\eea
In both cases, the   decay rate $\Gamma_2$ of this lighter singlet
is $\sim 45 H$, where $H$ is the Hubble expansion rate at
$T \sim M_2$, so leptogenesis 
would take place in the strong washout regime. 
In these limits the heavier singlet mass 
is found to be $M_3 \approx 2 \cdot 10^{15} ~{\rm GeV} $ 
in the hierarchical case and 
$M_3 \approx 6 \cdot 10^{14} ~{\rm GeV} $ in the inverse hierarchy.

\subsection{ complex case}
\label{complex}

 In this  section,
we check that the (unknown) phases of $W_L$ can always
be chosen  to compensate those
of $U$, so as to obtain negligeable imaginary
parts on the matrix $V_L$. We perform this
check for a hierarchical light neutrino spectrum. Recall that 
the matrix elements of 
$V_L$ control  the $\llg$ rates. If we could not
arrange their imaginary parts to be negligeable,
then our approximation of  neglecting phases would
not work.

 As  only two of the light neutrinos have masses,  only one Majorana phase 
is required in the lepton mixing  matrix $U$, so 
we set $\a =0$. The  $2 \times 2 $ unitary matrix  $W_L$ 
 can be written  
\beq 
W = e^{i\phi_g} \bma 1 & 0& 0 \\ 0 & c_{WL} e^{i \gamma/2} & - s_{WL} e^{-i\sigma/2} \\ 0 & s_{WL} e^{i \sigma/2} & c_{WL} e^{-i\gamma/2} \ema 
~~~.
\eeq
As in the previous section, we assume that
\beq
|V_{L3\mu} V_{L3e}^*|
= \left| \left( e^{\frac{1}{2}i(\gamma + \sigma + 2 \b -2 \delta)} 
c_{WL} s_{13} + c_{13} s_{12} s_{WL}  \right) \cdot \left( c_{WL} c_{13} s_{23} 
+ e^{\frac{1}{2}i(\gamma + \sigma +2 \b) } 
s_{WL} (c_{12} c{23} - e^{-i \delta} s_{12} s_{13} s_{23}) \right) \right| 
\eeq
 must be small to suppress
$\meg$,  from which we  want to determine the mixing angle 
$\theta_{WL}$. To obtain a value of the order of $10^{-3}$  
or even smaller, both  the real and  
imaginary parts of one of  the two parentheses must be small. 
The first parenthese is small for
\beq \label{eq:phase1} \begin{split} \frac{1}{2} (\gamma +\sigma + 2 \b - 2 \delta) \approx 0, \pm \pi, \pm 2\pi, ... \\
 \theta_{WL} \approx \arctan \left( \frac{s_{13}}{c_{13}s_{12}} \right) = 0.316 \end{split} ~~,
\eeq 
and the second is small for
\beq \label{eq:phase2} 
\begin{split} \frac{1}{2} 
( \gamma + \sigma + 2 \b ) 
& \approx \mbox{arccot}  
\left( \frac{\cos \delta - \frac{c_{12} c_{23}}
{s_{12}s_{13}s_{23}}}{\sin \delta} \right) 
= \mbox{arccot}  \left( \frac{\cos \delta - 11.2}{\sin \delta} \right) \\
\theta_{WL} & \approx - 0.77
\end{split} \eeq
We see that the condition of small $\mu \to e \gamma$ 
branching ratio  restricts  the possible phases.

An important question is  
whether  the predictions of the other two branching ratios 
depend strongly on the choice of the phases. 
We find  that  these two decay 
rates also only depend on the sum of $ \gamma + \sigma$. 
 If the phases vary in 
such a way that they still satisfy equation (\ref{eq:phase1}) 
or (\ref{eq:phase2}) and if $|V_{L3\mu} V_{L3e}^*|$ stays constant, we 
find that the other two branching ratios vary. 
But their overall variations over the whole parameter 
space are not bigger than a factor two. 
This factor two might be crucial for an actual 
determination of the branching ratios, but in the 
scope of our approximations this factor two will not 
influence our conclusions.

The phases also have an influence on the reconstructed Majorana masses. The first observation that we make is, that the Majorana mass eigenvalues become complex. However by one phase rotation we can absorb one phase in the ${\bf W}$ matrix. So finally we only have one complex mass eigenvalue. Finally we look at the absolute values of our Majorana masses and compare them to their value in the real case. We find that the variation does not exceed a factor four. This will also not have any influence on our conclusions.

In the following we will neglect phases in our reconstruction procedure. As we are not interested in $CP$ violation and the variations of the physical observables for different phases are rather small, we will place all our following analysis in a real case.


\section{Three singlets}
\label{3N}

This section studies the prospects for reconstruction
of  a seesaw model with  three heavy singlets $N_I$.
In this model, all the light neutrinos can be massive. 
We study the limit where the smallest light neutrino mass
$m_{\nu,min} \to 0$, to represent the experimental
situation where $m_{\nu,min} \ll \sqrt{\Delta m_{sol}^2}$
is unknown.  
This limit  could arise \cite{2RHN} when
 one of the singlets
is  very heavy, $1/M_3 \to 0$, so does not contribute
to the observed light masses, 
 or  when 
 one of the ${\bf Y}^\nu$ eigenvalues does
not contribute to the light masses,  $y_1 \to 0$.
We wish to know if such three singlet models
 can be distinguished from the two-singlet model of the
previous section. All phases are neglected (set to zero) in this section.

\subsection{One infinite mass eigenvalue}
\label{M3}

As regards the light neutrino mass matrix $[{\bf m}]$, 
 the  three singlet case with  $1/M_3 \to 0$
is   similar to the model of two singlets, 
because  $N_3$ makes no contribution to 
 $[{\bf m}]$ by construction. 
If, in addition, $N_3$ makes no contribution
 to the coefficients ${\bf X}_P$ of
the electromagnetic dipole operator
(see eqn (\ref{masina})),
 then
$N_3$ is  decoupled and    the model
looks like a two-singlet model.

The aim here is  to consider 
the situation where $N_3$ did contribute
to ${\bf X}_P$, but not to $[{\bf m}]$. In
a supersymmetric scenario,
 this situation could arise in the
very narrow window where  $M_3< M_X$
(recall that $M_X \lsim m_{pl}$ is the scale at which
the soft masses appear), but $M_3$
is irrelevant for $[{\bf m}]$. 
It is clear that when  the mass $M_3$  exceeds
the scale $M_X$,  then the heaviest singlet
is not present to contribute to loop corrections
to the soft masses. The combination of
Yukawas   ${\bf Y}^\nu {\bf Y}^{\nu \dagger}$ in
eqn (\ref{Xmfv}), would  therefore  be replaced by
\beq
\label{masina}
{\Big [}{\bf  Y}^\nu {\bf Y}^{\nu \dagger}{\Big]}_{M_3 \to \infty} = 
{\Big [}V_L^\dagger {\bf D}_\nu V_R P_{12} V_R^\dagger  
{\bf D}_\nu V_L{\Big ]}
~~~~~, ~~~~ P_{12} \equiv
\left[
\begin{array}{ccc}
1 & 0 & 0 \\
0 & 1 & 0 \\
0 & 0 & 0
\end{array}
\right] ~~,
\eeq
where $P_{12}$ is the projector onto the two lighter 
 singlets, in the $N_I$ mass eigenstate basis.
 It is straightforward to check that
this matrix has two non-zero eigenvalues. If it
is identified  as a $2 \times 2 $  matrix 
${\bf Y}^\nu {\bf Y}^{\nu \dagger}$,
then,  in combination with the light neutrino mass matrix
$[{\bf m}]$, it allows to reconstruct the 
seesaw model  of the two lighter singlets. The
heaviest is fully decoupled and irrelevant.  

\begin{figure}[ht]
\begin{center}
\epsfig{file=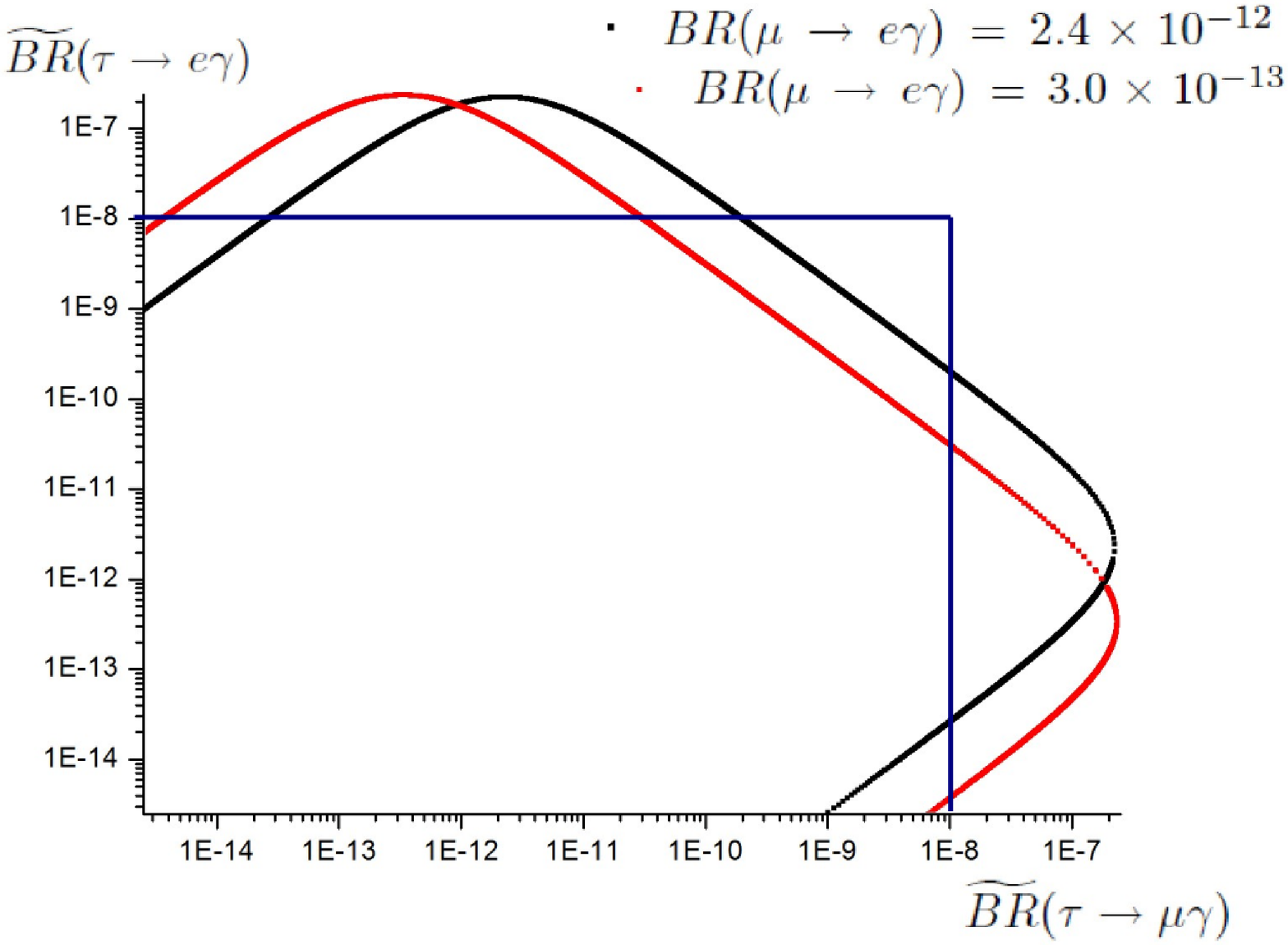, height=8cm,width=13cm}
\end{center}
\caption{ Branching ratios for $\tmg$
and $\teg$ (normalised to leptonic decays as
in eqn (\ref{radiative_decay})) in  the 
hypothetical cases
that $\meg$ was observed,
with 
 $BR(\meg) = 2.4 \times 10^{-12}$
(upper rising-towards-the-left  black  line), or
 $BR(\meg) = 3.0 \times 10^{-13}$
(lower rising-towards-the-left  red line). The correlation
arises  from the assumptions discussed
after eqn (\ref{rapports}). 
The $\widetilde{BR}$s are estimated
with eqn (\ref{raddec3}). The estimated
sensitivity of  superB factories
 is outside the (blue) rectangle.
This plot can also be interpreted as an
upper bound on the $\widetilde{BR}$s 
which are consistent with a  $\meg$
bound:
the allowed region is to the left
and below 
 the diagonal line sloping-up-to-the-left.
The horizontal and vertical cutoffs
are  uncertain because
they are determined by our estimate of
$C_\nu^X$.
\label{fig:BRratios}}
\end{figure}

For the remainder of this subsection,  we assume that
the Yukawa matrix ${\bf Y^\nu}$  has three non zero eigenvalues,
 so that  $W_L$ (or $V_L$) and $V_R$ are  $3 \times 3$ 
matrices, 
described by three mixing angles.
If  $V_L ^\dagger$ is written in the form of the
leptonic mixing  matrix $U$,  then 
with the approximation (\ref{approxhier}),
 two of the angles  could
be determined from two ratios of branching ratios :
\beq 
\label{rapports}
\frac{\widetilde{BR}(\meg)}{\widetilde{BR}(\teg)} = 
\frac{ |V_{L 3\mu}|^2}{|V_{L 3 \tau}|^2} =
\frac{s_{VL23} ^2}{c_{VL23} ^2}
~~~~~~~~
\frac{\widetilde{BR}(\meg)}{\widetilde{BR}(\tmg)} = 
\frac{ |V_{L 3e}|^2}{|V_{L 3 \tau}|^2} =
\frac{s_{VL13} ^2}{c_{VL23} ^2 c_{VL13} ^2}  ~~.
 \eeq
That means that measuring one of these
ratios of branching ratios, does not predict
the other (This is different from the two singlet
model of the previous section). 
However,  if
 $|V_{L3 \mu} V_{L3 e}^*| $
is fixed by an observation of $\meg$,
then measuring one ratio
of eqn (\ref{rapports})
determines the other 
--- this is illustrated 
in figure \ref{fig:BRratios}, where the
upper diagonal (black) corresponds
to  $BR(\meg) = 2.4 \times 10^{-12}$,
and  the lower diagonal (red) line
would correspond to 
 $BR(\meg) = 3.0 \times 10^{-13}$
(The branching ratios are estimated
via eqn (\ref{raddec3})).
 This correlation
is not a consequence of  the
supersymmetric  seesaw model;
 rather, it  follows from assuming that:
\ben
\item
the coefficient-matrix ${\bf X}_L$ of 
the dipole operator contains
one lepton flavour changing matrix
as in eqn  (\ref{Xmfv}),
which is dominated by its largest
eigenvalue, and,
\item  that
the small branching ratio of
$\meg$ imposes an approximate
zero in ${\bf X}_L$.
\een
 So if the prediction
of figure \ref{fig:BRratios} was verified, it would
be consistent with the three-singlet
seesaw, as well as any other MFV-like
model satisfying the above two assumptions. 
And if some other ratios of branching
ratios were measured,  it could likely
be accomodated in a supersymmetric 
three-singlet  seesaw model by relaxing the above
assumptions.

To reconstruct the singlet masses
with  equation(\ref{DM-1}), requires 
 the neutrino mass matrix ${\bf m_\nu}$
and  the whole ${\bf P}$ matrix. 
However,
in the approximation (\ref{approxhier}),
neither the ${\bf Y}^\nu$ eigenvalues nor
 the third mixing angle of $V_L$  can be 
determined from rare decays   (degenerate sneutrinos 
\cite{GH,snus} could give additional
information). Our  reconstructed
singlet masses therefore  depend on
the unknown $\theta_{VL 12}, y_1$ and $y_2$.
We can  estimate ranges for these parameters
which give $M_1 \sim 10^{10}$ GeV, as would be suitable
for thermal leptogenesis. 

\begin{figure}[h]
\centering
\includegraphics[height=7cm]{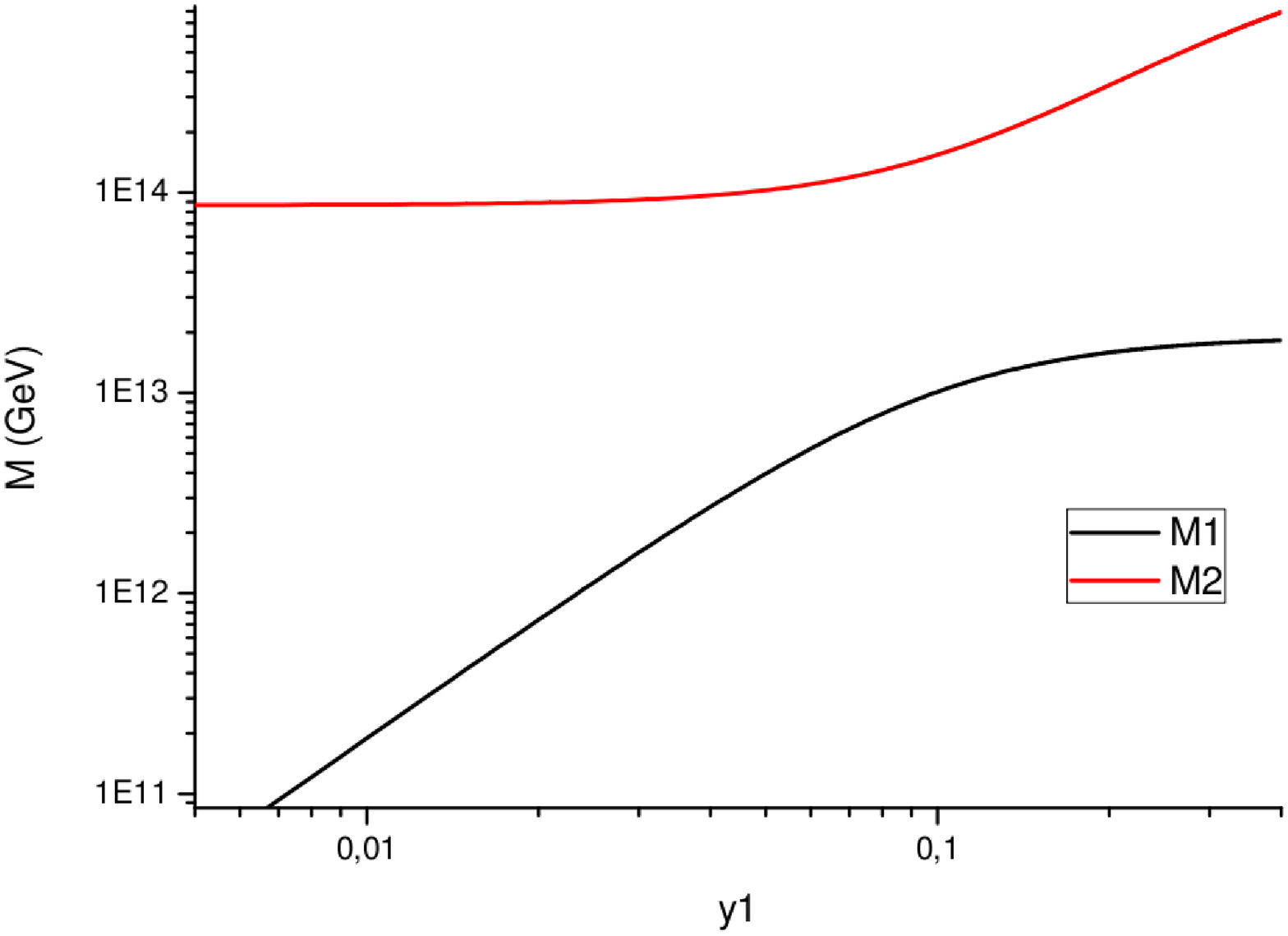}
\caption{General dependence, in the three-singlet
model where $1/M_3 \to 0$,  of  the singlet  masses 
on the smallest Yukawa eigenvalue $y_1$. In
this plot, $y_2 = .01$ and $y_3 = 1$.
The angles in $W_L$ were taken 
similar to those of $U$: $\theta_{WL13}  = .18$,
 $\theta_{WL12}  = .6$, and
 $\theta_{WL23}  = .6$.
Different values of the mixing angles
can shift the normalisation of the axes.
\label{fig:dependence}}
\end{figure}

The inverse mass matrix of the singlets 
is given in equation(\ref{DM-1}), as a function of
the ``known'' doublet parameters.
When one eigenvalue  of a $3 \times 3$ matrix  ${\bf M}$ vanishes,
the two remaining eigenvalues are given by \cite{DLR}
\beq 
M_1 , M_2 = \frac{1}{2} \left[ 
\mbox{Tr}[{\bf M}] 
 \pm \sqrt{
(\mbox{Tr}[{\bf M}])^2 -4({ M}_{11} { M}_{22} + {M}_{11}{M}_{33} 
+ M_{22} M_{33} - M_{23}^2 - {M}_{12} ^2 -{M}_{13} ^2)} \right] 
\eeq
We find that for generic choices of
the mixing angles in $V_L$,  
the   singlet  masses $M_1$ and $M_2$  
have the form  $ \sim v^2 y_i^2/\overline{m}$, as plotted in figure
\ref{fig:dependence}, where
$\overline{m} = \sqrt{ \Delta m^2_{sol}}
+\sqrt{ \Delta m^2_{atm}}$.  This  is
unsurprising, since the light masses
corresponding to eqn (\ref{m2}) are
much more degenerate than the 
squared eigenvalues of ${\bf Y}^\nu$. 
So a given  $M_I$  is usually   controlled by a
single $y_i^2$, and 
the  singlet mixing angles in $V_R$
are generically  small. That is, $V_R$ usually  can be written
as a rotation matrix with small mixing angles 
multiplied by either the identity matrix   or a permutation. 
We find that we can
always  choose Yukawa eigenvalues  
to  obtain $M_1 \sim 10^{10} ~{\rm GeV}$, as 
suggested for thermal leptogenesis.
 This tells us  that even if  all rare decays were measured,
 the lightest singlet mass is not determined,
so whether it has a value suitable
for leptogenesis  is not known.

To reproduce,  in this model, the  two singlet case 
of section \ref{2N},
 can  be achieved by setting $\theta_{WL12}$ 
and $\theta_{WL13}$ in $W_L$ to zero. 
Then   $y_1$ becomes  
irrelevant for the reconstruction, 
$\theta_{WL23}$ plays the role of $\theta_{WL}$ in
 the two singlet model. In this limit, $M_1$ 
 depends quadratic on $y_2$. 

An interesting question is whether  we can 
distinguish between  the  two or three singlet
models  by measuring rare decay rates? Not unambiguously. 
If  the  measured branching ratios disagree with the predictions 
of the two singlet see-saw model, it can be excluded. 
However, if the branching ratios agree with the  
two singlet predictions, we cannot exclude
the three singlet model, which can fit  every constellation of 
branching ratios that is possible in the two singlet model.

Consider now the case where
 the smallest neutrino mass is small, but non-zero. 
The Majorana mass eigenvalue $M_3$ is no longer infinite,
 but should remain larger than the other two mass eigenvalues $M_{1,2}$. 
However, it is not transparent  to analytically
determine the values of the smallest mass, as a function of
other parameters, for which this condition
is satisfied. 
A more stringent condition that ensures the validity of 
our reconstruction is that, in equation (\ref{DM-1}),
 for each component of $\bf M^{-1}$, the 
effect of $m_1$ is small compared to the effects 
of the other two mass eigenvalues (normal hierarchy):
\beq \label{cond}
  [{\bf D_{\nu}}^{-1} { W_L}]_{I1} m_1 
[{ W}_L^T {\bf D}_{\nu}^{-1} ]_{1J} << [{\bf D_{\nu}}^{-1} { W_L}]_{Ik} m_k 
[{ W}_L^T {\bf D}_{\nu}^{-1} ]_{kJ} \,\,\,\, \mbox{ where } k = 2,3 \mbox{ and } I,J = 1,2,3
\eeq
Those conditions 
imply that the reconstruction 
done in this section is valid  if $m_1 << m_k$, except 
when $W_L$ is   almost 
a diagonal or permutation matrix. 
In this 
special case,  the contributions of $m_{2,3}$ are multiplied 
by 
a small mixing angle or by the cosine 
of a almost $\pi/2$ angle, whereas the contribution 
of $m_1$ is multiplied by almost one, so the
reeconstruction is  valid for
\beq 
\frac{m_1}{m_k} << \theta_{WLi}^2 \mbox{ or } <<\left(  \theta_{WLi}- \frac{\pi}{2} \right)^2
\eeq

\subsection{One zero Yukawa eigenvalue}
\label{y1}

In this section, we briefly discuss the case
where 
the lightest neutrino mass  vanishes,  
$m_{\nu min} \rightarrow 0$, because  
one Yukawa 
eigenvalue, which we take to be 
$y_1$, goes to zero. 

Equation (\ref{Dnu}) can be rewritten:
\beq \label{Dm}     W_L {\bf D}_m W_L ^T 
= {\bf D}_\nu V_R {\bf D_M}^{-1} V_R ^T {\bf D}_\nu v_u ^2 
\eeq
where ${\bf D_\nu}$ has only two non zero entries 
on the diagonal and therefore the right hand side of this 
equation has only a $ 2 \times 2$ sub matrix with 
non-zero entries. Furthermore, we know that the 
doublet neutrino mass eigenstates in ${\bf D}_m $ do not 
couple to the vanishing first Yukawa eigenvalue $y_1$. 
Therefore the two angles $\theta_{W12}$ and $\theta_{W13} $ 
are identically zero, and 
a $2 \times 2$ singlet inverse mass matrix can be reconstructed
using eqn (\ref{DM-1}).   
However, ${\bf D}_M^{-1}$ on the right hand side of
eqn (\ref{Dm})  still 
contains the three singlet masses. 
Three equations do not  
determine three mass eigenvalues and 
three mixing angles,
so the whole high energy Lagrangian
cannot be reconstructed.

This limit reduces to the case of two right 
handed neutrinos. The right hand side of equation 
(\ref{Dm}) can be rewritten in terms of two 
"effective" mass eigenvalues and one 
"effective" mixing angle. 
Of course the effective values are functions of the 
three initial mass eigenvalues and the three mixing angles, 
but we can only determine the effective values. 
So we cannot observe any difference between 
the two models unless we can measure 
the operator ${\bf [d]}$. 
With this operator we can distinguish 
between the two models, but we cannot 
perform a complete reconstruction, as discussed in  the following section.

\section{More singlets and the inverse seesaw}
\label{4N}

In this section, we consider the prospects for
testing, in principle, the seesaw model we reconstructed.  
We suppose an ideal world where the matrices 
$ {\bf P},
[{\bf m}] $ and
$[{\bf d}] $ could be exactly determined. Then from
 $ {\bf P} $ and 
$[{\bf m}] $, one can compute  ${\bf d}$, and
compare its elements
to their measured values.

As a first example,  consider a  model 
with 
three singlets,  and  $m_1 =  y_1 = 0$.
To be concrete, we 
take a normal hierarchy for the light neutrino
masses; a similar reasoning
applies for  the inverse
hierarchy, but in a different   
two-dimensional  subspace of doublet flavour  space.

In the singlet  basis where ${\bf Y^\nu}^\dagger{\bf  Y^\nu}$ is
diagonal (with eigenvalues $\{ 0, y_2, y_3 \}$), 
we can reconstruct a $2\times 2$ submatrix, which we call
 ${\bf {\widetilde M}}^{-1}$,  of the 
``true'' $3 \times 3$   matrix  
\beq
{\bf M}^{-1} = 
\left[
\begin{array}{ccc} 
M^{-1}_{11} & M^{-1}_{12} &M^{-1}_{13}\\
M^{-1}_{21} & \widetilde{M}^{-1}_{22} 
& \widetilde{M}^{-1}_{33}\\
M^{-1}_{31} &  \widetilde{M}^{-1}_{32} & \widetilde{M}^{-1}_{33}\\
\end{array}\right]~~~,
\eeq
We imagine this matrix is 
 reconstructed from  $ {\bf P}$ and 
$[{\bf m}] $, as was discussed in section
\ref{3N}: one obtains the Yukawa eigenvalues
(two non-zero) and the $3 \times 3$ rotation matrix
$V_L$  from   $ {\bf P}$. This gives the
$2 \times 2$  matrix $W_L$, which can be
substituted into eqn (\ref{DM-1}) whose left-hand-side is
$\widetilde{{\bf M}}^{-1}$.

The matrix  ${\bf {\widetilde M}}^{-1}
{\bf {\widetilde M}}^{*-1}$,
together with $W_L$ and ${\bf D}_\nu$, 
can  combined to give a ``prediction'' for $[{\bf d}]$,
expressed in the neutrino
mass basis:
\beq
\label{dpred}
[{\bf d}] {\Big |}_{pred} =
W_L^\dagger {\bf D}_\nu{\bf {\widetilde M}}^{-1}
{\bf {\widetilde M}}^{*-1} 
 {\bf D}_\nu W_L
\eeq
which differs from the ``true'' expression for $[{\bf d}]
=W_L^\dagger {\bf D}_\nu
 {\bf  M}^{-1}
{\bf  M}^{*-1}
 {\bf D}_\nu W_L $,
\beq
 [{\bf  d}]  {\Big |}_{true}
=
W_L^\dagger {\bf D}_\nu
\left(
 {\bf {\widetilde M}}^{-1}
{\bf {\widetilde M}}^{*-1} +
\left[
\begin{array}{cc}
 |M^{-1}_{12}|^2 &  M^{-1}_{12}  M^{*-1}_{13} \\
 M^{*-1}_{12}  M^{-1}_{13} &  |M^{-1}_{13}|^2
\end{array}
\right] \right)
 {\bf D}_\nu W_L
\eeq 
which would correspond to the measured value,
in our ideal world. 
It is clear that the reconstructed 
and true matrices differ, and the determinant of
the difference is zero.  We could repeat
this exercise with four singlets, and
find that the determinant of the
difference matrix is non-zero.  So 
in principle, in this best of ideal
worlds, it is possible to find traces of
additional singlets, even when their
(Yukawa) interactions with  the Standard
Model are irrelevant to LFV and neutrino
masses.

Notice that since $y_1 = 0$, we not can
get access to $[{\bf M}^{-1}]_{11}$, and therefore
can not determine the mass of the  singlet
$N_1^c$. So a complete reconstruction
is not possible.  This is as expected, because in the limit
where $[{\bf M}^{-1}]_{1i} \to 0$,for $i \neq 1$, the singlet $N_1^c$
is decoupled from the Standard Model.

Finally, consider an ``inverse seesaw''  model, with
two singlets $N_I$ of lepton number -1,  and two singlets
$S_I$ of lepton number 1, who have no
Yukawa couplings but a small majorana mass matrix
{\boldmath $\mu$}$_{IJ}$:
\beq
{\cal L}_{inverse} =  
 y^e_{{\alpha}} \overline{\ell}^{\alpha}  H^*_d  e_R^{\alpha}+ 
   \overline{\ell}^{\alpha}   H^*_u  [  {\bf Y^\nu}^*]_{\alpha I} N_I  + 
 {M}_I  \overline{S}_I N_I +  \frac{\mu_{IJ}}{2}  \overline{S^c}_I S_J+ h.c.
\label{seesawinverse}
\eeq 
This is a special case of a four-singlet model with
only two  non-zero eigenvalues in  ${\bf Y^\nu}$,
and   we just  argued  that a generic such  four-singlet model
could  not be reconstructed even in the best of all worlds.
Nonetheless, 
the  supersymmetric inverse seesaw {\it is} reconstructable. From
$ {\bf P}$   and $ [{\bf d}]$, which are defined
as in the ordinary seesaw in eqns (\ref{P}) and (\ref{opBGJ}),
one can obtain  ${\bf Y^\nu}$ and ${\bf M}$. Then from 
the effective light neutrino mass matrix
\beq
[{\bf m}] =  {\bf {\bf Y^\nu}} {\bf M}^{-1}
\mu \,  {\bf M}^{-1}
 {\bf Y^\nu}^{T} v_u^2 ~~~ ,~~~~  
\eeq
one can obtain {\boldmath $[\mu]$}. This could suggest that
the supersymmetric inverse seesaw could fit  any
observations of neutrino mass, mixing and non-unitarity,
as well as   $\ell_\a \to \ell_\b \gamma$ \cite{IMP,Shaban}.
An independent determination of the  singlet masses(at a collider),
could test this model.


\section{Summary and Discussion }
\label{disc}

We considered  simple toy 
supersymmetric seesaws (the lightest neutrino
was taken  massless) with 2,3 and 4 heavy singlets,
and usually real parameters. We 
were interested in the properties of
a model that allow it to be reconstructed
in principle, so we  studied
the prospects of reconstructing the high-scale
parameters of these models from  observations
at or below a TeV. We imagine that such observations
could give us access to  three matrices
in doublet lepton flavour space (see
eqns (\ref{P}), (\ref{m_nu_eff}), and (\ref{opBGJ})) :
$$ {\bf P} = V_L^\dagger D^2_{\nu} V_L~~~  ,~~~  
[{\bf m}] =  {\bf {\bf Y^\nu}} {\bf M}^{-1} {\bf Y^\nu}^{T} v_u^2 ~~~ ,~~~~  
[{\bf d}] =  {\bf Y^\nu}{\bf M}^{-1} {\bf M}^{-1 *} {\bf Y^\nu}^\dagger 
$$ 
It is a straightforward exercise in linear algebra to
show that, if  $ {\bf Y^\nu}$ is an  invertible $3 \times 3 $
matrix,  then  from any pair
of the above three matrices, 
a $3 \times3$ matrix ${\bf M}$ can be obtained.
This gives the masses and mixing angles of three
singlets $\{ N_I \}$. Therefore,  measuring
the third of the above three matrices could
test the model. 

This programme to test the seesaw suffers
from  a serious practical problem, which is that 
$  {\bf P},  {\bf m}$ and $ {\bf d}$ cannot be measured
exactly, if at all. We attempted to circumvent this
problem by neglecting  phases, and by using ``reasonable''
theoretical expectations (such as that the
$ {\bf Y^\nu}$ eigenvalues are hierarchical)
for unknown parameters. 
We also take the lightest neutrino mass to be zero.

Then we studied the question of ``principle'',
whether it is possible to determine the
number of singlets participating in the seesaw,
and reconstruct their properties. 
\bit
\item Since the lightest neutrino was taken massless,
only two singlets are requireed to reproduce
the light neutrino mass matrix.  As shown by Ibarra
\cite{2RHN}, the seesaw model with  
two singlets, and two eigenvalues for $ {\bf Y^\nu}$,
 is testable.   In particular, as discussed in
section \ref{2N} it predicts a discrete
set of possible rates for $\tlg$, given the rate
for $\meg$.  If the $BR(\meg)$ is taken at the current bound of 
$2.4 \times 10^{-12}$, then for a normal hierarchy of
light neutrino masses, one of the  $BR(\tlg) \gsim 10^{-8}$ 
should be visible at a super-B factory (for
susy parameters as described in the 
Appendix, giving rate expectations 
as in  eqn (\ref{raddec3})). In the case of
an inverse hierarchy, a similar situation
could arise, or --- unfortunately---
both $\tlg$ decays could be out of the reach
of super-B factories. 

To reconstruct the singlet sector of this model, 
we assume the largest ${\bf Y}_\nu$ eigenvalue to
be 1, and fix the smaller by setting the lighter
singlet mass $\sim 10^{10}$ GeV, for thermal leptogenesis. 

\item The 
three singlet model can have a massless light neutrino
in two limits:  either one
of the singlets is very heavy ($M_3^{-1} \to 0$),
or 
there are only two non-zero eigenvalues to
in $ {\bf Y^\nu}$ (so $y^\nu_1 \to 0$).
\bit
\item
The first case, where the
$3 \times 3$   $ {\bf Y^\nu}$ is invertible,
is   in principle a reconstructable model,
allowing to obtain the mixing angles and 
masses of the three singlets.
As discussed in section \ref{3N} 
it differs from the two-singlet model, 
in that the three angles of $V_L$ are
independent. This means that  knowing the rate
 for one Lepton Flavour Violating decay,
such as $\meg$, does not determine
the other two.  Instead, for hierarchical
 $ {\bf Y^\nu}$ eigenvalues, knowing
two lepton-flavour-violating branching ratios, allows to
predict the third. 

However, as discussed in
section \ref{3N}, this  case is not a very
realistic limit: if $M_3$ is sufficiently
large that it contributes negligeably to the light
neutrino masses, it is unlikely to
be  present  in loop corrections
to the sneutrino masses. This means
that the Yukawa couplings of $N_3$
would not contribute to ${\bf P}$,
which would  therefore  effectively
have only two non-zero eigenvalues
(see eqn (\ref{masina})).

\item The second case, where the smallest of
the three  Yukawa eigenvalues is negligeable,
is equivalent to the  two-singlet case,
except that the $2 \times 2$ singlet
inverse mass 
matrix  that one reconstructs using
a $2 \times 2 $ $ {\bf Y^\nu}$ matrix
in  eqn (\ref{DM-1}), 
is only a $2 \times 2 $ submatrix
of the full inverse mass
matrix ${\bf M}^{-1}$. As discussed
in section \ref{4N}, models which all
have two  non-zero $ {\bf Y^\nu}$ eigenvalues,
but have  two, three or four singlets,
are in principle distinguishable because 
they give  different predictions for
${\bf d}$. However, it is not possible to
fully reconstruct the singlet sector.
This is straightforard to see in the limit
 where the additional singlets  are  decoupled,
because they interact  only via
the vanishing Yukawa eigenvalue.

\eit
\item
Four singlet models are in general
not reconstructible --- that is,
the singlet masses cannot be obtained.
However,  if the singlet
mass matrix is of the  ``inverse seesaw''
type, it    is   reconstructible, in principle,
from the sneutrino mass matrix (${\bf P}$)
the light neutrino mass matrix,  and
 the operator ${\bf d}$ given in eqn
(\ref{opBGJ}). 
\eit

Almost any  observed rates  for $\llg$
can be compatible with 
the three-or-more  singlet
supersymmetric  seesaw \cite{DI1,ISconf}.
The precise relation between 
the $\llg$  rates,  and the seesaw parameters
depends on various aspects  and scales of the 
soft-supersymmetry-breaking mechanism. To circumvent
the (not very well delimited) uncertainties 
resulting from
these  ``soft'' assumptions, 
we estimated the $\llg$ rates using
a basic Minimal-Flavour-Violation-like
parametrisation (eqn (\ref{Xmfv})), which could
give a ``lowest order'' decription of the
supersymmetric seesaw --- and of several
other models. 

The coefficient ${\bf X}_L$ of the electromagnetic
dipole operator (see eqn (\ref{dipole})) is 
a matrix with indices in doublet and singlet
charged lepton flavour spaces. 
We assumed  that its 
flavour-changing elements were  proportional
to ${\bf Y}^{e\dagger} \hat{{\bf P}}$, where
${\hat{\bf P}}$ is a new ``spurion'', or
matrix in doublet lepton  flavour space.
A first approximation to
the supersymmetric seesaw is to take
$\hat{{\bf P}} = {\bf Y}^{\nu}{\bf Y}^{\nu \dagger}$.
We further assumed that the eigenvalues of 
$ \hat{{\bf P}}$ were hierarchical, such
that the largest contributions to
flavour changing matrix elements were always
 mediated by the  largest eigenvalue. 
Finally, we assumed that the upper
bound on $\meg$ imposes an approximate
zero in $\hat{{\bf P}}$.
These assumptions, which could roughly
describe  the three-singlet supersymmetric
seesaw, predict a relation between 
the $\tmg$ and $\teg$ rates 
 (see
eqn (\ref{rapports}) and figure \ref{fig:BRratios}).

The context of this study, was
to explore the prospects of  ``reconstructing''
the neutrino mass generation mechanism
from coefficients of the effective Lagrangian.
We focussed on supersymmetric seesaw models,
because  seesaws  have many features 
which facilitate reconstruction. 
They   contain New Physics at tree level, with 
 well-separated  mass scales.
This means that the heavy propagators
can be approximated as $1/m^2$ ---
which would be less straightforward,
for instance, if the New Physics appeared in loops.
In addition, all the  new 
 interactions and masses  of the
seesaw are flavoured and bilinear.  
This means that coefficients in the
effective Lagrangian are constructed by
matrix multiplication, so by
combining ``suitable'' coefficients, one
can solve for all the new matrices.  
This would be less simple, for instance,
with R-parity violating trilinear couplings.
Nonetheless, these
nice features of the seesaw are
not  sufficient to ensure reconstructability, 
because  we obviously  cannot reconstruct New Physics in 
the limit where it decouples from us.
It would be interesting to know what features
are neccessary for a reconstructable model.

\subsection*{Acknowledgements}

We thank Sebastien Descotes-Genon, Benjamin Guiot,
 and Christopher Smith for 
useful conversations. 

\section*{Appendix}

Our model is supersymmetric, so 
 soft supersymmetry-breaking 
terms  for the ``sparticles'' must be
added to the Lagrangian
obtained from the superpotential
of eqn (\ref{W}).  These renormalisable
terms are interesting for seesaw reconstruction, because
loop processes can imprint traces
of the high-scale theory upon them.   A
``Minimal Flavour Violation\cite{dAGIS}-like'' 
expectation for the mass-squared matrix of the slepton doublets
$\{ \widetilde{L}_\a \}$ could  be
of the form 
\beq
\label{softmfv2}
[\widetilde{{\bf m}}^2]  =  C^m_I {\bf I} 
+ C^m_{e} {\bf Y^e} {\bf Y^e}^\dagger + 
C^m_{\nu} {\bf Y^\nu} {\bf Y^\nu}^\dagger
\eeq
because, in the high-scale theory
described by the superpotential
of eqn (\ref{W}),
 the  two spurions
linking the  {\boldmath $3$} and  {\boldmath$\bar{3}$} 
of the doublet lepton space are
${\bf Y^e} {\bf Y^e}^\dagger$ and ${\bf Y^\nu} {\bf Y^\nu}^\dagger$.
In the charged lepton mass eigenstate basis,
 ${\bf Y^\nu} {\bf Y^\nu}^\dagger$ generally
gives   off-diagonal
elements of  $[\widetilde{{\bf m}}^2]$,
which would  induce  Lepton Flavour Violating (LFV) processes
\cite{FrancescaBM},
such as  
$\ell_\a \to \ell_\b \gamma$.

In a model for supersymmetry-breaking and flavour,
the soft supersymmetry breaking parameters can be
calculated, which  gives a prediction
for  the coefficients $C^m_x$, and
for corrections\cite{LPR} to eqn  (\ref{softmfv2}). 
An estimate for 
$C^m_\nu$ can be obtained in gravity-mediated supersymmetry-breaking
scenarios, because  the soft breaking parameters
are present at scales $\gg M$,
where loops involving
Higgses(higgsinos) and singlet (s)neutrinos give
small  contributions to the  doublet
slepton masses. 
These are approximately
proportional to 
$ {\bf Y^\nu} {\bf Y^\nu}^\dagger$,
and are resummed by the renormalisation group
equations.  Assuming that,
at some high scale $M_X \gg M_I$, 
the model contains universal
 soft terms of the form
\beq
\mathcal{L}_{soft} =  \tilde{m}_0^2  
\tilde{L}_\a^{\dagger} \tilde{L}_\a 
+ a_0( y^e_{\alpha} {\widetilde{L}}^{\alpha} \cdot  H_d  
\widetilde{E^c}_{\alpha} +
{\bf Y^\nu}_{\alpha I} \widetilde{L}^{\alpha} \cdot  H_u 
\widetilde{N^c}_I)  + h.c. ~~,
\label{brisSUSY}
\eeq
then, in the leading log approximation,
 the 
slepton doublet  soft masses
 $\widetilde{m}^2_{\alpha \beta}$
receive   contributions
  \cite{Masina,FP} of order
\bea \label{m_soft} 
\Delta \widetilde{m}^2_{L\alpha \beta} &\simeq& 
-\frac{3 m_0^2 + a_0^2}{16 \pi^2} 
 \left[{\bf Y^\nu}  
\log \left( \frac{M M^{\dagger}}{M_X^2} \right) {\bf Y^\nu}^{\dagger}\right]_{\a \b}
\eea
where $\alpha, \beta$ are indices in the
charged lepton mass basis.  
If one assumes  that 
 all the
singlets
 can be  decoupled at the same scale $M_{av}$,
and that $\tilde{m}_0 = a_0$,  
then one can  approximate  \footnote{
Analytic reconstruction is possible
without this assumption as shown
in section 3.2.3 of \cite{ybook}.}
\beq \label{approxn+1}
 \left[{\bf Y^\nu}  
\log \left( \frac{M M^{\dagger}}{M_X^2} \right) {\bf Y^\nu}^{\dagger}\right]_{\a \b}
 \to  
\left[ {\bf Y^\nu}   {\bf Y^\nu}^{\dagger}\right]_{\a \b}
 \log\left( \frac{M_{av}^2}
{M_X^2}\right) 
\eeq
which gives
\beq
\label{Cnu}
C^m_\nu \simeq  -\frac{ m_0^2}{4 \pi^2} 
\eeq
where we approximate the $\log \to 1$.

Lepton flavour violating 
radiative
decays,  $\ell_\a \to \ell_\b \gamma$,
  proceed
via the operator of eqn (\ref{dipole}) 
\beq
e X_{L\a \b} \overline{e_\a} \sigma ^{\mu \nu} P_L e_\b F_{\mu \nu}
+
e X_{R\a \b} \overline{e_\a} \sigma ^{\mu \nu} P_R e_\b F_{\mu \nu}
~~~,
\label{dipole2}
\eeq
at a rate given by eqn (\ref{radiative_decay}) 
\bea \label{radiative_decay2}
\widetilde{BR} (l_{\alpha} \to l_{\beta} ~ \gamma) = 
\frac{\Gamma (l_{\alpha} \to l_{\beta} ~ \gamma)}
{\Gamma(\ell_\alpha \rightarrow \ell_\beta \nu_{\alpha} \bar{\nu}_{\beta})} 
&   =  & 
\alpha m_\alpha^3 (|X_{L \a \b}|^2 + |X_{R \a \b}|^2) \frac{192 \pi^3} 
{G_F^2 m_\alpha^5} ~~.
\eea
For a given weak-scale SUSY spectrum, the coefficients $X_{P\a\b}$
can be calculated \cite{HisanoPLB,HisanoPRD}
 as a function of  masses and mixing
angles.

For  $\a = \b = \mu$, the dipole operator (\ref{dipole})
also contributes to the anomalous magnetic moment of
the muon 
\beq
\frac{g-2}{2} \equiv  a_\mu = 4 {\rm Re}  \{ X_{ L \mu \mu} \} m_\mu
\eeq
whose experimental determination \cite{g-2}
differs from the SM prediction by several 
standard deviations \cite{Davier:2010nc}.
If this anomaly   $\delta a_\mu  \simeq 29 \times 10^{-10}  $
\cite{Davier:2010nc} is dominantly
induced by the same   New Physics diagram as  
$\ell_\a \to \ell_\b \gamma$ (for instance,
a chargino-slepton loop ---see \cite{GT}
for intuitive estimates of the various
diagrams in the mass insertion approximation),
then \cite{Moroi,GT}
\beq
\label{number}
\delta a_\mu  \simeq
\frac{g_2^2 m_\mu^2}{32 \pi^2 \tilde{m}^2_{SUSY}} \tan \b
\eeq
and 
it is possible to ``normalise'' 
 \cite{HisanoTobe}
the amplitude for   $\ell_\a \to \ell_\b \gamma$  
by  $\delta a_\mu  $:
\bea
X_{L \alpha \beta} &   \sim  & X_{L \alpha \a}
\frac{\widetilde{m}^2_{\alpha \beta}}{\widetilde{m}^2}
\sim 
\frac{\delta a_\mu}{4 m_\mu^2} m_{\alpha}  
\frac{\widetilde{m}^2_{\alpha \beta}}{\widetilde{m}^2} ~~~.
\eea
For $ \widetilde{m}^2 = m_0^2$, 
$\widetilde{m}^2_{\alpha \beta} $ from eqn (\ref{m_soft}), 
and $X_{L \alpha \beta}$ of the form (\ref{Xmfv}):
\beq
{\bf X}_{L\a \b}  =
 m_\a C^X_{\nu}  {\bf Y^\nu} {\bf Y^\nu}^\dagger ~~~, \a \neq \b
\eeq 
this gives 
\bea
\label{CnuX}
C^X_{\nu} &   \sim  & 
\frac{\delta a_\mu}{16 \pi^2 m^2_\mu}    ~~~,
\eea
and
\bea \label{raddec3}
\widetilde{BR} (l_{\alpha} \to l_{\beta} ~ \gamma)
&   =  & 
\frac{3  \alpha  |\delta a_\mu|^2} 
{ 4 \pi  m_\mu^4 G_F^2 } 
{\Big |} [{\bf Y^\nu} {\bf Y^\nu}^\dagger]_{\a \b} {\Big |}^2
\simeq 10^{-6}{\Big |}  [{\bf Y^\nu} {\bf Y^\nu}^\dagger]_{\a \b} {\Big |}^2
~~.
\eea

As is well-known \cite{ISconf,toutlemonde},  
the upper bound  $\widetilde{BR}(\meg) \leq 2.4 \times 10^{-12}$ \cite{MEG}
 restricts
$[{\bf Y^\nu} {\bf Y^\nu}^\dagger]_{\mu e}   \lsim 
   2  \times 10^{-3}$,
which  implies, with the hierarchical
assumption for ${\bf Y}^\nu$ of eqn
(\ref{approxhier}), 
that there is a small entry in the mixing matrix $V_L$:
\bea
\label{small2}
y_3^2 |V_{L3\mu} V_{L3 e}^*| &  \lsim &  2 \times 10^{-3} ~~~.
\eea
From current bounds \cite{tmgexp}
 on $\tmg$ and $\teg$,
 similar approximations give 
$ | [{\bf Y^\nu} {\bf Y^\nu}^\dagger]_{\tau \ell} |  \lsim  0.5$.

Notice that for $\tan \b = 10$, saturating
the $(g-2)_\mu$ anomaly via  eqn (\ref{number})
gives $\tilde{m}_{SUSY} \sim  225$ GeV, 
a scale where
colliders do not seem to find a supersymmetric menagerie. 
The estimate (\ref{raddec3}) scales with various
parameters as
\bea \label{raddec4}
\widetilde{BR} (l_{\alpha} \to l_{\beta} ~ \gamma)
&   \simeq  & 
\simeq 10^{-6}
\left\{  \, 
\left(\frac{\tan \b}{10} \right)^2
\left(\frac{225 {\rm GeV}}{\tilde{m}_{SUSY}} \right)^4
\right\}
{\Big |}  [{\bf Y^\nu} {\bf Y^\nu}^\dagger]_{\a \b} {\Big |}^2
~~,
\eea
so changing a parameter by a factor of 2,
can change the rare decay rates by
an order  of magnitude.  The ``approximate zero''
of eqn (\ref{small}), and the expectations
which follow from it,  relies on the factor
in curly brackets  in (\ref{raddec4}) being
larger than $ 10^{-3} $.

\end{document}